\documentclass[
dvips,
amsmath,
amssymb,
aps,
showpacs,
preprintnumbers,
superscriptaddress,%
]{revtex4-1}

\usepackage{graphicx}
\usepackage{dcolumn}
\usepackage{bm}
\usepackage{color}

\begin{document}

\title{ The Study of Ratio Algorithm for Radiographic Imaging with Cosmic-ray Muons}
\newcommand{\ustc}{University of Science and Technology of China, Hefei 230026, China}
\newcommand{\ahu}{AnHui University, Hefei 23061, China}

\author{Liu Cheng-Ming}\affiliation{\ustc}
\author{Wen Qun-Gang}\email{qungang@ahu.edu.cn}\affiliation{\ahu}
\author{Zhang Zhi-Yong}\email{zhzhy@ustc.edu.cn}\affiliation{\ustc}
\author{Huang Guang-Shun}\email{hgs@ustc.edu.cn}\affiliation{\ustc}
\date{\today}
\begin{abstract}
  The paper\cite{Nature2003}  presented a novel muon radiography technique which exploits the multiple Coulomb scattering of these particles for nondestructive inspection without the use of artificial radiation.
  In this paper, a new kind algorithm named Ratio Algorithm was proposed for imaging with the test object.
  The experimentally produced cosmic-ray muons radiographs was reconstructed.
  The more important, the reconstruction was made using data from only 793 muons.

\end{abstract}
\maketitle

\section{Introduction}
Cosmic-ray muons imaging is a new developed technique in material detecting.
The muon is a kind of secondary particle that comes from the extensive atmosphere shower of high-energy cosmic rays, the average energy scale is about $3\sim 4\rm GeV$, the high energy and strong penetration of muon guaranteed the Coulomb scattering inside materials.
After knowing the changing of incident track and ejection track, the scatterpoint could be defined, so the tracks should be precisely measured.

In order to obtain the tracks, the Micro mesh gaseous structure (Micromegas) was used for tracks' measurement in this experiment.
Micromegas detector is a new type of gaseous detector based on planar electrodes, it has the abilities of high counting rate, high gain, high position resolution and Micromegas is cheaper comparing to the solid detectors.
Therefore, it is a good choice for muon imaging experiment.

Due to the low flux of muon at the sea level, which is just $\rm 10,000\ m^{-2} min^{-1}$ , so it is very important to use the imaging algorithm to analyze the data collected, based on the experiment, a new algorithm name ratio algorithm was proposed in this paper.

\section{The Ratio Algorithm}
In this paper, a new kind algorithm named Ratio Algorithm was proposed for imaging with the test object.
Considering the randomness of the muon's movement in material, the distribution of multiple Coulomb scattering approximates a Gaussian distribution\cite{PRD_Theory}.
The many small interactions add up to yield an angular deviation that roughly follows as Eq.(\ref{eq:gaus_1}),
\begin{equation}
  \label{eq:gaus_1}
  \frac{dN}{d\theta}=\frac{1}{\sqrt{2\pi}\sigma}\exp\big(-\frac{\theta^2}{2\sigma^2}\big)
\end{equation}
with the width, $\sigma$, related to the scattering material through its radiation length, $L_{rad}$, as follows:
\begin{equation}
  \label{eq:gaus_2}
  \sigma = \frac{13.6\mathrm{MeV}}{\beta cp}\cdot\frac{L}{L_{rad}}\Big[1+0.038\ln\big(\frac{L}{L_{rad}}\big)\Big]
\end{equation}
where $p$ is the particle's momentum in MeV$\cdot c^{-1}$ and $\beta c$ is its velocity$^2$.
For a certain $L$, the radiation length decreases rapidly as the atomic number $Z$ of a material increases, and $\sigma$ increases accordingly.
\begin{figure}[thb]
  \centering
\includegraphics[scale=0.9]{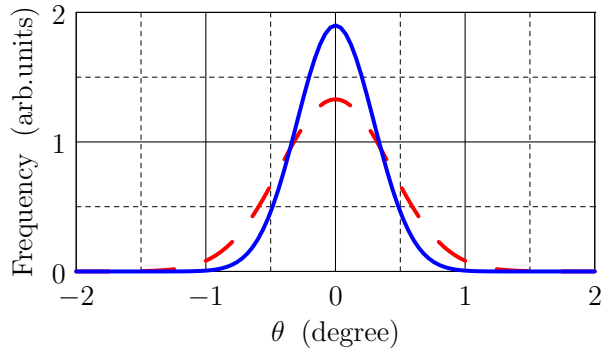}
\caption{The behavior of muons passing through the material.}
\label{fig:gaus}
\end{figure}
The Fig.\ref{fig:gaus} showed the behavior of muons passing through the material.
The blue line was the angle ditribution of muons passing through the low atomic number $Z$ material, the red dashed line represented the high $Z$ material.
In an experiment, all the events detected were counted as $A_0$, and events in centain angle range (See Fig.\ref{fig:gaus}, for example $|\theta| < 0.6$) were counted as $A_c$.
The ratio $R$ value was defined as
\begin{equation}
  \label{eq:Ratio}
  R = \frac{A_c}{A_0}
\end{equation}
Obviously, for materials with different $Z$ value, the higher $Z$ material would derive lower $R$ value.
In this way, the different material could be distinguished by $R$ value.
What’s more, the $R$ value could also be linked to the $Z$ value if the tracks were precise enough.

\section{Experiment And Image Reconstruction}
\begin{figure}[htb]
  \centering
  \includegraphics[scale=0.9]{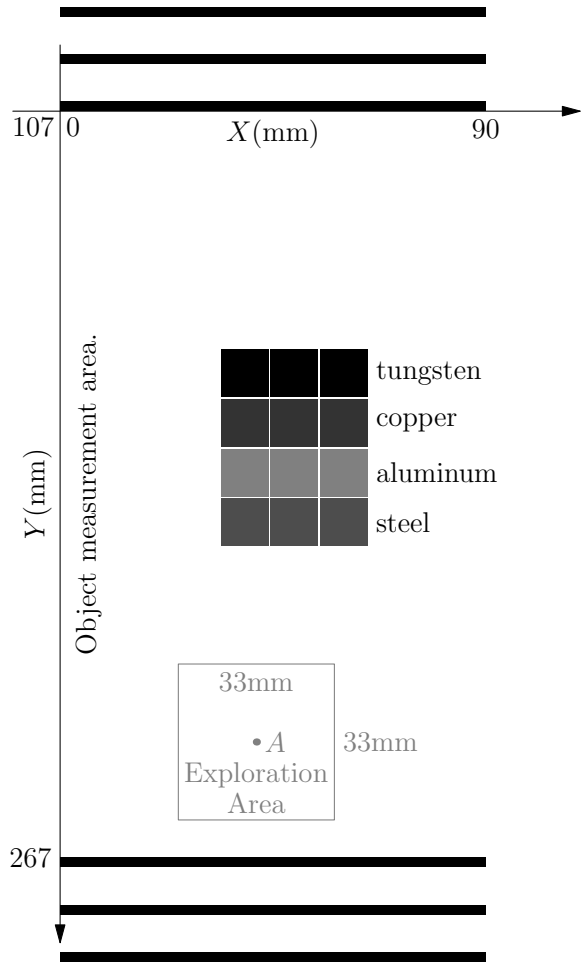}  
  \caption{Sketch of the experimental setup.}
  \label{fig:setup}
\end{figure}

To test the validity of the Ratio algorithm, we constructed a small experimental apparatus built with a set of six  Micromegas detectors.
These detectors have been made using the thermal bonding technique which was researched and developed by the University of Science and Technology of China (USTC).
The sensitive area of all six detectors was $90\mathrm{mm}\times 90\mathrm{mm}$ and were designed as one dimension readout.
Three groups of detectors were placed above an object region to record the tracks of incident muons, and three groups were placed below to record the scattered tracks, according to the configuration shown in Fig.\ref{fig:setup}.
Two plastic scintillators were used for giving the trigger for the MPD plugins in the VME.
The data acquisited by MPD plugins through the APV25 chips were recorded by the computer and then used for analyzing.
%
The spatial resolution of the Micromegas detectors used is better than 200$\mu m$.
  
The material used for detecting were twelve metal rods included four kinds of metal listed as steel, aluminum, copper and tungsten.
They are all $\rm 10mm\times 10mm\times 100mm$ and were bundled as a bigger rod with the cross section of $\rm 30mm\times 40mm$ (Shown in the Fig.\ref{fig:setup}).

Muons pass through three sensitive detectors above the object measurement area, providing incoming angle and position.
They then pass through the object measurement area and are scattered to an extent dependent upon the material through which they pass.
The position and angle of scattered particles are measured in other three detectors.
Each detector measures particle position in one orthogonal coordinate.
Using the ratio algorithm to  reconstruct the material, we need to get a ratio for every point in the object measurement area.
The ratio of one point can be getten as:
\begin{itemize}
\item Selecte a random point in object measurement area.
  Such as $A$ point in Fig.\ref{fig:setup}.
\item Select an area centered on this point to be exploration area.
  The exploration area is $\rm 33mm\times 33mm$ in Fig.\ref{fig:setup}.
\item All events pass through the exploration area were counted as $A_0$, and events which scatter angle in centain range ($|\theta|<1.6^{\circ}$ in this work) were counted as $A_c$.
\item Use Eq.(\ref{eq:Ratio}) to get the ratio $R$ for this point.
\end{itemize}
We randomly selected 500,000 points in the object measurement area, and calculated the $R$ for every point.
Then a quantity field $R(x, y)$ of the object measurement area could be obtained experimentally. 
Using computer software to visualize the experimental data, we can get the cosmic-ray muon radiographs.

\section{Results And Discussion}
\begin{figure}[tbh]
  \centering
  \includegraphics[scale=0.3]{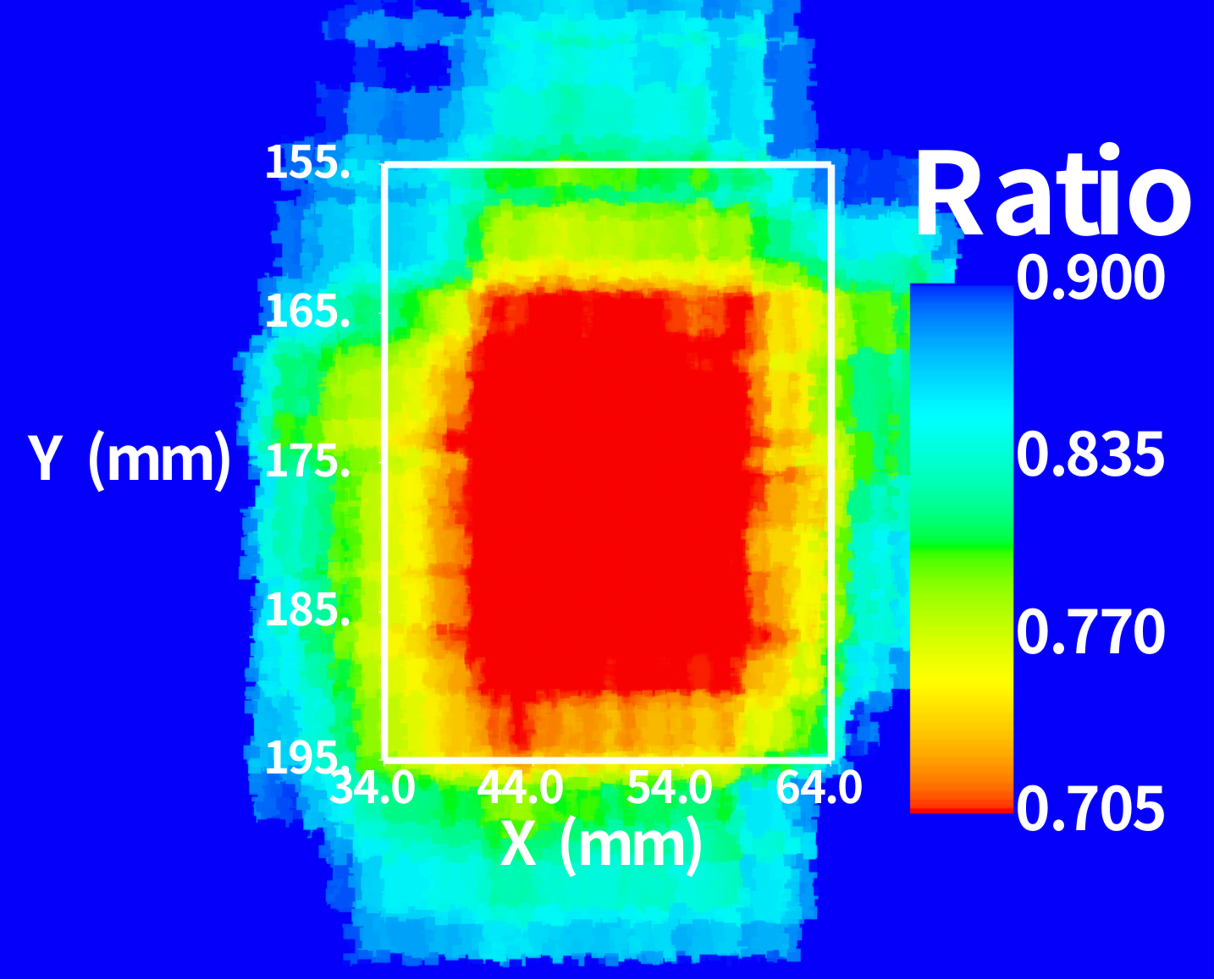}  
  \caption{Experimentally produced cosmic-ray muon radiographs.}
  \label{fig:image}
\end{figure}

The experimentally produced cosmic-ray muons radiographs was shown in Fig.\ref{fig:image}.
The rectangular box with coordinates shows the the location and size of the test object.
These points with ratio less than $0.705$ are roughly formed into a rectangle.
The reconstructed rectangle is smaller than the actual test object, but the reconstructed position is consistent with the actual position.
These points with ratio more than $0.90$ represent air.
The ratio from $0.705$ to $0.90$ mainly occurs in the transition region between the test object (high-$Z$) and air (low-$Z$).
It is important to point out that the reconstruction was made using data from only 793 muons.
This shows that this algorithm will be very useful for rapid detection.

\end{document}